# Modified Lagrangian Formulation of Gear Tooth Crack Analysis using Combined Approach of Variable Mode Decomposition (VMD) – Time Synchronous Averaging (TSA)

**Subrata Mukherjee[1,2,@], Vikash Kumar[2] and Somnath Sarangi[2]**

**Thapar Institute of Engineering and Technology, Patiala, Punjab 147004, India**
**Indian Institute of Technology Patna, Bihar 801106, India**

**[@]Correspoding Author: subrata.re.me@gmail.com**

**Abstract.**

This paper discusses the possible observation of an integrated gear tooth crack analysis procedure that employs the combined approach of variable mode decomposition (VMD) and time synchronous averaging (TSA) based on the coupled electromechanical gearbox (CEMG) system. This paper also incorporates the modified Lagrangian formulation to model the CEMG system by considering Rayleigh's dissipative potential. An analytical improved time-varying mesh stiffness (IAM-TVMS) with different levels of gear tooth crack depts is also incorporated into the CEMG system to inspect the influence of cracks on the system's dynamic behavior. Dynamic responses of the CEMG system with different tooth crack levels have been used for further investigations. For the first time, the integrated approach of variable mode decomposition (VMD) and time-synchronous averaging (TSA) has been presented to analyze the dynamic behaviour of CEMG systems at the different gear tooth cracks have been experienced as non-stationary and complex vibration signals with noise. Based on the integrated approach of VMD-TSA, two types of nonlinear features, i.e., Lyapunov Exponent (LE) and Correlation Dimension (CD), were calculated to predict the level of chaotic vibration and complexity of the CEMG system at the different levels of gear tooth cracks. Also, the LE and CD are used as chaotic behaviour features to predict the gear tooth crack propagation level. The results of the proposed approach show significant improvements in the gear tooth crack analysis based on the chaotic features. Also, this is one of the first attempts to study the CEMG system using chaotic features based on the combined approach of VMD-TSA.

**Keywords.** IAM-TVMS, coupled electromechanical model, modified Lagrangian formulation, gear tooth cracks, dynamic response, variable mode decomposition (VMD), time synchronous averaging (TSA), Lyapunov Exponent (LE), Correlation Dimension (CD), chaotic vibration, and chaotic complexity.

## 1. Introduction

Vibration signal-based gearbox health monitoring under non-stationary conditions has been an essential requirement for modern machine maintenance in various industries such as automotive, aerospace, mining, and so on. Gears experienced severe functioning conditions and unwarranted service loads in numerous industrial applications, leading to fault growth. In the gearbox, 60% of failure occurs due to various gear faults, such as fatigue cracks in the gear tooth [1-5]. But due to the non-stationary conditions, the gearbox is subjected to an extensive range of interference in vibration signal extraction, making it very difficult to diagnose fatigue cracks early [1-5].

Mathematical or physical modelling of gearbox systems, including gear tooth faults, has been widely helpful for gearbox fault diagnosis [1-4]. With the help of mathematical modelling, researchers can find the exact nature of the vibration signal in the presence of tooth faults [1-4]. Also, simulated vibration response has well-defined taring datasets that can be easily used in machine learning-based fault diagnostics [5]. In gearbox modelling, one of the most important critical issues is the formulation and evaluation of time-varying mesh stiffness (TVMS) with different tooth faults, which can significantly affect the dynamic response of the gearbox [1-4].

On this note, two well-known TVMS formulation methods, namely the analytical and finite element (FE) methods, significantly contributed to gearbox modeling. Analytical approaches are more straightforward and have higher computational efficiency [1-4]. In contrast, the FE modelling approach is more appropriate if the extended tooth contact occurs outside the theoretical contact line, especially if the gear crack is larger [1-4].

At the same time, various signal processing techniques are easily computable to extract the optimal vibration signal in the presence of nonstationarities [5-12]. Conservative procedures such as TSA [7,11,12] are usually applicable for minimizing the effects of noise and the asynchronous part of the vibration signal. Similarly, frequency domain approaches like FFT deliver information about the dominant frequency components of vibration signals [5-12].

However, these techniques are limited to handling vibration signals' nonstationarities and instantaneous frequency components [5-12]. Therefore, time-frequency methods such as EMD, STFT, WT, and WVM were developed to optimize the above shortcomings [5,6]. These techniques can handle complex and non-stationary vibration signals [5,6]. Currently, a variable mode decomposition (VMD) technique is used with the development of time-frequency methods to recognize disguised fault transients and filter out the non-stationary, which provides concurrent modes extraction with an improved center frequency for each mode of vibration signal [7-10].

Even though there are multiple gear tooth cracks, models are available to predict gear tooth faults at an early stage [1-4]. However, a unified gear tooth crack model with varying degrees of tooth crack propagation is required to control better multiple parameters of gear tooth cracks at an early stage, such as chaotic complexity and chaotic features. Furthermore, it is critical to develop an integrated and unified mathematical model coupled electromechanical gearbox (CEMG) system that can provide a mechanical and electrical perspective of gear tooth crack propagation by measuring chaotic complexity and features.

Therefore, an inclusive methodology has been proposed to analyze the gear tooth crack of a coupled electromechanical gearbox (CEMG) system based on the complex and non-stationary dynamic response to noise that can be implemented in real-time industrial sectors. This study provides a comprehensive explanation for efficiently diagnosing early gear cracks under nonstationarities, impacting operating machines' coupled electromechanical gearbox (CEMG) system. The proposed methodology of this paper has been presented in Figure 1. The novel contributions that have been made in this present study are as follows:

- A unified mathematical model of a coupled electromechanical gearbox system has been incorporated, which was developed in [3] based on the modified Lagrangian formulation by considering Rayleigh's dissipative potential.

- An analytically improved time-varying mesh stiffness (IAM-TVMS) model [3,4] also has been incorporated with the developed mathematical model of coupled electromechanical gearbox system.

- An integrated signal preprocessing technique has been proposed and presented by combining VMD with TSA that calculates the Intrinsic Mode Functions (IMFs) to filter out nonstationarities from the complex and non-stationary dynamic response to noise. Later, these IMFs also have been preprocessed with TSA to minimize the effects of noise and asynchronous components.

- A list of nonlinear features, i.e., Lyapunov Exponent (LE) and Correlation Dimension (CD), are calculated from each IMF-TSA to validate and enumerate some critical observations of the proposed method by measuring the level of chaotic vibration and chaotic complexity in the VMD-TSA dynamic response.

The article is structured as follows: Section 2 discusses the necessary theoretical background. The developed Lagrangian formulation of the coupled electromechanical system, including IAM-TVMS and an improved gear crack model, has been presented in Section 3. Section 4 discusses the proposed integrated VMD-TSA signal processing methodology. Section 5 discusses the proposed method's results and effectiveness, and Section 6 reports the significant conclusions drawn from the present work.

## 2. Theoretical Background

### *2.1 Variable Mode Decomposition (VMD)*

VMD has been proposed by Dragomiretskiy et al. [9] to decompose a multi-mono-component signal collaborative of band-limited intrinsic mode functions (IMFs), which have applied numerous fault diagnosis approaches [7,8,10].

Then, the VMD algorithm was demonstrated as follows:

Step 1: Initialize $\{\hat{o}_k^1\}, \{\hat{e}_k^1\}, \hat{L}^n, n \leftarrow 0.$ (1)

Step 2: The value of $o_k$, $e_k$, and $L$ is updated according to the following mathematical formula:

$$\hat{o}_k^{n+1} \leftarrow \frac{\hat{f}(e) - \sum_{i<k} \hat{o}_i^{n+1}(e) - \sum_{i>k} \hat{o}_i^{n+1}(e) + \frac{\hat{L}^n(e)}{2}}{1 + 2\alpha\left(e - e_k^n\right)^2} \quad (2)$$

$$e_k^{n+1} \leftarrow \frac{\int_0^{+\infty} e\left|\hat{o}_i^{n+1}(e)\right|^2 de}{\int_0^{+\infty} \left|\hat{o}_i^{n+1}(e)\right|^2 de} \quad (3)$$

$$\hat{L}^{n+1}(e) \leftarrow \hat{L}^n(e) + \tau[\hat{Z}(e) - \sum_K \hat{o}_i^{n+1}(e) \quad (4)$$

Step 3. Repeat step 2 until the function satisfies the condition, $\sum_K \frac{\left\|\hat{o}_k^{n+1} - \hat{o}_k^n\right\|_2^2}{\left\|\hat{o}_k^n\right\|_2^2} < \epsilon$, where $\epsilon$ is the specified precision claim.

### *2.2 Time Synchronous Averaging (TSA).*

TSA, commonly implemented as a time-domain averaging technique to extract periodic components, is one of the strongest and most operative signal processing methods for extracting periodic signals from a composite signal applied to rotating machinery [7,11,12].

If we consider the raw vibration signal to be a continuous signal, the linear averaging operation is given by [8-10]:

$$Z(t) = \frac{1}{L} \sum_{n=0}^{L-1} Z(t - nV) \quad (5)$$

where V is the rotational period, and L is the number of averages.

### *2.3 Correlation Dimension (CD)*

The correlation dimension can be used for self-similarity quantification of vibration signals by measuring the chaotic complexity [13,16,17]. A more considerable value means higher complexity and less similarity. Mathematically, the CD is defined as:

$$\mathrm{CD} = \lim_{r \to 0} \frac{log \sum_{i=1}^{Nr} Z_i^2}{logr} \quad (6)$$

In this, *N(r)* is the amount of d-dimensional cells of size r, $Z_i = \frac{X_i}{X}$ is the probability where X & $X_i$ , are the total amount of points in the set and set-in hypercube *i*.

### *2.4 Lyapunov Exponent (LE)*

The Lyapunov exponent measures the exponential divergence of two initial adjacent trajectories in phase space [13-15,17].

A negative value indicates convergence, while a positive one demonstrates variation and chaos. Whenever a chaotic system is in progress from a preliminary condition in phase space by the radius of $A_0$ and after the time duration *t*, this radius is changed to:

$$A = A_0 x^{\varphi t} \quad (7)$$

In equation 7, the value of $\varphi$ is known as Lyapunov Exponent

## 3. Proposed Methodology

In this section, the description of the proposed method has been demonstrated. Fig. 1 shows the flowchart of the proposed methodology. The following points describe the proposed model that has been developed:

1. Firstly, an analytically improved TVMS model and improved tooth crack model were incorporated with an eight-degree freedom coupled electromechanical gearbox system developed based on the modified Lagrangian formulation considering Rayleigh's dissipative potential.

2. Then, the developed model was investigated based on the two speed-load conditions, i.e., 25Hz-25lb and 25Hz-50lb, and four gear tooth conditions, i.e., healthy, 20% tooth cracks, 40% tooth cracks and 60% tooth cracks with additions of -10dB and 10dB SNR.

3. After that, the VMD was performed on all cases and filtered the nonstationarities from the dynamic response, which was later used for TSA preprocessing to remove the effects of noise and asynchronous components from the VMD (IMFs/VMFs) vibration signal.

4. LE and CD were calculated from each VMF-TSA signal of all described conditions using the m = 3 & d =1, and the essential observations of the proposed method were discussed.

## 4. Mathematical Modelling of Coupled Electromechanical Gearbox

### *4.1 Modified Langrange Formulation of Coupled Electromechanical Gearbox Model*

This work incorporates an eight-degree-of-freedom coupled electromechanical gearbox (CEMG) system based on the modified Langrange formulation previously derived and presented in [3]. The Langrange formulation takes Rayleigh's dissipative potential to establish the mathematical model. Figure 2 shows the representation of the physical interpretation of the developed CEMG system. The equations of motion (EoM) for the CEMG system are written as [3]:

$$V_{as} = R_s I_{as} + L_{ss}\frac{dI_{as}}{dt} - \frac{1}{2}L_{ms}\frac{dI_{bs}}{dt} - \frac{1}{2}L_{ms}\frac{dI_{cs}}{dt} + L_{ms}\cos\theta_r\frac{dI_{ar}}{dt} + L_{ms}\cos(\theta_r + 2\pi/3)\frac{dI_{br}}{dt} + L_{ms}\cos(\theta_r - 2\pi/3)\frac{dI_{cr}}{dt} - L_{ms}[(I_{ar}\sin(\theta_r) + I_{br}\sin(\theta_r + 2\pi/3) + I_{cr}\sin(\theta_r - 2\pi/3)]\frac{d\theta_r}{dt} \quad (8)$$

$$V_{bs} = R_s I_{bs} + L_{ss}\frac{dI_{bs}}{dt} - \frac{1}{2}L_{ms}\frac{dI_{as}}{dt} - \frac{1}{2}L_{ms}\frac{dI_{cs}}{dt} + L_{ms}\cos\theta_r\frac{dI_{br}}{dt} + L_{ms}\cos(\theta_r - 2\pi/3)\frac{dI_{ar}}{dt} + L_{ms}\cos(\theta_r + 2\pi/3)\frac{dI_{cr}}{dt} - L_{ms}[(I_{br}\sin(\theta_r) + I_{cr}\sin(\theta_r + 2\pi/3) + I_{ar}\sin(\theta_r - 2\pi/3)]\frac{d\theta_r}{dt} \quad (9)$$

$$V_{cs} = R_s I_{cs} + L_{ss}\frac{dI_{cs}}{dt} - \frac{1}{2}L_{ms}\frac{dI_{as}}{dt} - \frac{1}{2}L_{ms}\frac{dI_{bs}}{dt} + L_{ms}\cos\theta_r\frac{dI_{cr}}{dt} + L_{ms}\cos(\theta_r - 2\pi/3)\frac{dI_{br}}{dt} + L_{ms}\cos(\theta_r + 2\pi/3)\frac{dI_{ar}}{dt} - L_{ms}[(I_{cr}\sin(\theta_r) + I_{ar}\sin(\theta_r + 2\pi/3) + I_{br}\sin(\theta_r - 2\pi/3)]\frac{d\theta_r}{dt} \quad (10)$$

$$R_s I_{ar} + L_{ss}\frac{dI_{ar}}{dt} - \frac{1}{2}L_{ms}\frac{dI_{br}}{dt} - \frac{1}{2}L_{ms}\frac{dI_{cr}}{dt} + L_{ms}\cos\theta_r\frac{dI_{as}}{dt} + L_{ms}\cos(\theta_r + 2\pi/3)\frac{dI_{bs}}{dt} + L_{ms}\cos(\theta_r - 2\pi/3)\frac{dI_{cs}}{dt} - L_{ms}[(I_{as}\sin(\theta_r) + I_{bs}\sin(\theta_r + 2\pi/3) + I_{cs}\sin(\theta_r - 2\pi/3)]\frac{d\theta_r}{dt} = 0 \quad (11)$$

$$R_s I_{br} + L_{ss}\frac{dI_{br}}{dt} - \frac{1}{2}L_{ms}\frac{dI_{ar}}{dt} - \frac{1}{2}L_{ms}\frac{dI_{cr}}{dt} + L_{ms}\cos\theta_r\frac{dI_{bs}}{dt} + L_{ms}\cos(\theta_r - 2\pi/3)\frac{dI_{as}}{dt} + L_{ms}\cos(\theta_r + 2\pi/3)\frac{dI_{cs}}{dt} - L_{ms}[(I_{bs}\sin(\theta_r) + I_{cs}\sin(\theta_r + 2\pi/3) + I_{as}\sin(\theta_r - 2\pi/3)]\frac{d\theta_r}{dt} = 0 \quad (12)$$

$$R_s I_{cr} + L_{ss}\frac{dI_{cr}}{dt} - \frac{1}{2}L_{ms}\frac{dI_{ar}}{dt} - \frac{1}{2}L_{ms}\frac{dI_{br}}{dt} + L_{ms}\cos\theta_r\frac{dI_{cs}}{dt} + L_{ms}\cos(\theta_r - 2\pi/3)\frac{dI_{bs}}{dt} + L_{ms}\cos(\theta_r + 2\pi/3)\frac{dI_{as}}{dt} - L_{ms}[(I_{cs}\sin(\theta_r) + I_{as}\sin(\theta_r + 2\pi/3) + I_{bs}\sin(\theta_r - 2\pi/3)]\frac{d\theta_r}{dt} = 0 \quad (13)$$

With the help of kinetic energy, the electromagnetic torque of the system can be calculated [3] :

$$T_e = \frac{\partial K.E}{\partial \theta_r} \quad (14)$$

$$T_e = -L_{ms}[(I_{as}I_{ar} + I_{bs}I_{br} + I_{cs}I_{cr})\sin\theta_r + (I_{as}I_{cr} + I_{bs}I_{ar} + I_{cs}I_{br})\sin\left(\theta_r - \frac{2\pi}{3}\right) + (I_{as}I_{br} + I_{bs}I_{cr} + I_{cs}I_{ar})\sin(\theta_r + 2\pi/3)] \quad (15)$$

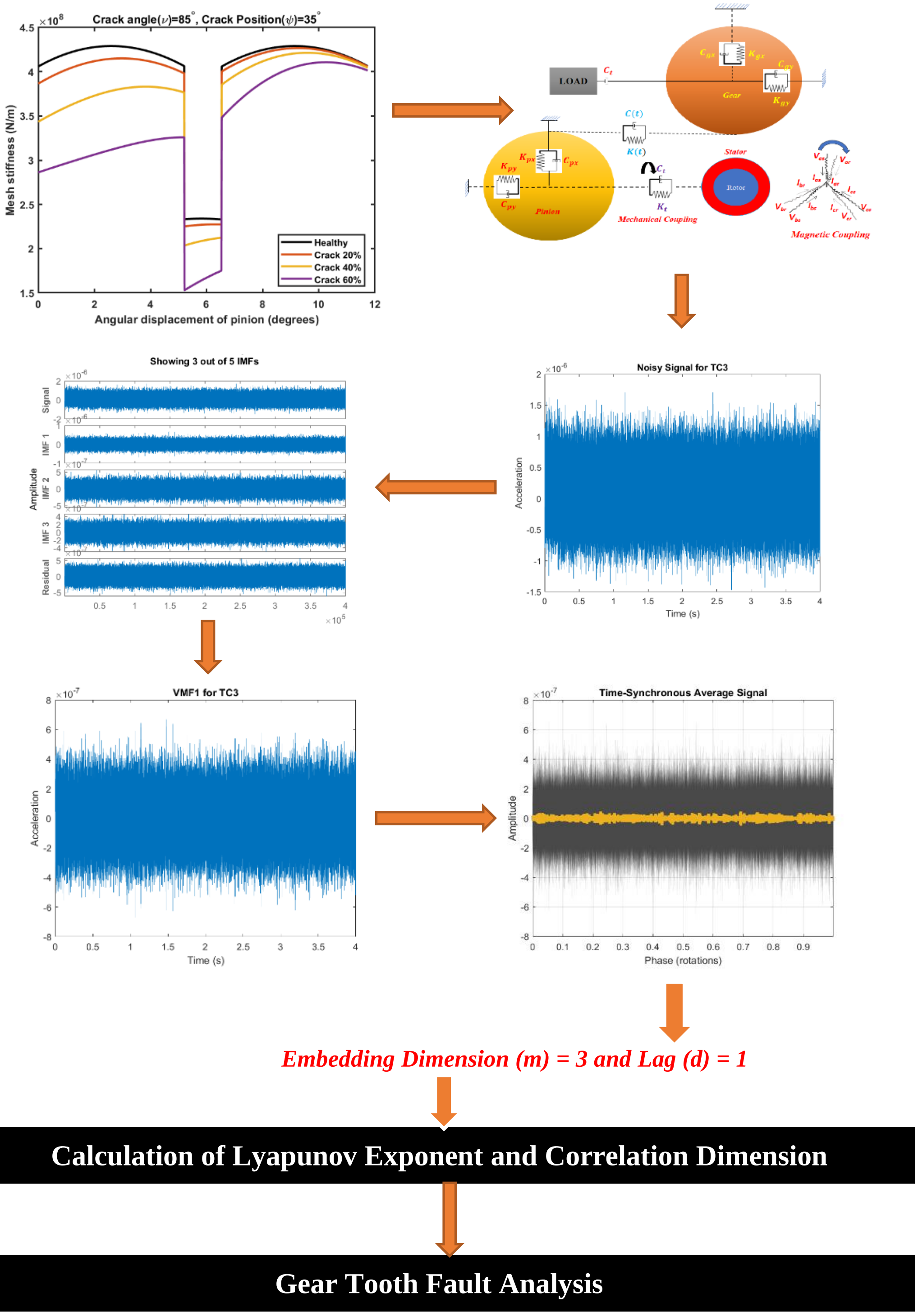

**Figure 1** Flowchart of the proposed methodology for gear tooth crack analysis using VMD-TSA

For the presence of derived input torque (15) in the EoM, any tooth crack faults that arise in the dynamic portion of the CEMG system may be reflected over the input current response, which is the function of the three phases' current signal [3].

$$m_p \frac{d^2 x_p}{dt^2} = -K_{xp} x_p - C_{xp} \frac{dx_p}{dt} + F_p \tag{16}$$

$$m_g \frac{d^2 x_g}{dt^2} = -K_{xg} x_g - C_{xg} \frac{dx_g}{dt} + F_g \tag{17}$$

$$m_p \frac{d^2 y_p}{dt^2} = -K_{yp} y_p - C_{yp} \frac{dy_p}{dt} - N \tag{18}$$

$$m_g \frac{d^2 y_g}{dt^2} = -K_{yg} y_g - C_{yg} \frac{dy_g}{dt} + N \tag{19}$$

$$i_p \frac{d^2 \theta_p}{dt^2} = r_p N - K_t(\theta_P - \theta_r) - C_t(\omega_P - \omega_r) + M_p \tag{20}$$

$$i_g \frac{d^2 \theta_g}{dt^2} = -\ r_g\ N +\ \ M_g -\ \ T_L - B_v \omega_g \tag{21}$$

$$i_m \frac{d^2 \theta_r}{dt^2} = -K_t(\theta_r - \theta_p) - C_t(\omega_r - \omega_P) - B_v \omega_r + T_e \tag{22}$$

In the above equations,

$$N = K(t)[(y_p - y_g) - (r_p\theta_p - r_g\theta_g)] + C(t)[\frac{d(y_p - y_g)}{dt} - \frac{d(r_p\theta_p - r_g\theta_g)}{dt}] \tag{23}$$

On this note, the total mesh stiffness and gear mesh damping coefficient are named *K(t)* and *C(t)*, respectively. Hence, *C(t)* is calculated by:

$$C(t) = 2 \times \zeta \times \sqrt{K(t)(m_p \times m_g)/(m_p + m_g)} \tag{24}$$

$M_p$ and $M_g$, is the moment due to friction force $F_p$ and $F_g$, respectively. Fig. 2 shows the physical model of the CEMG system with the used parameters to derive the EoM [3].

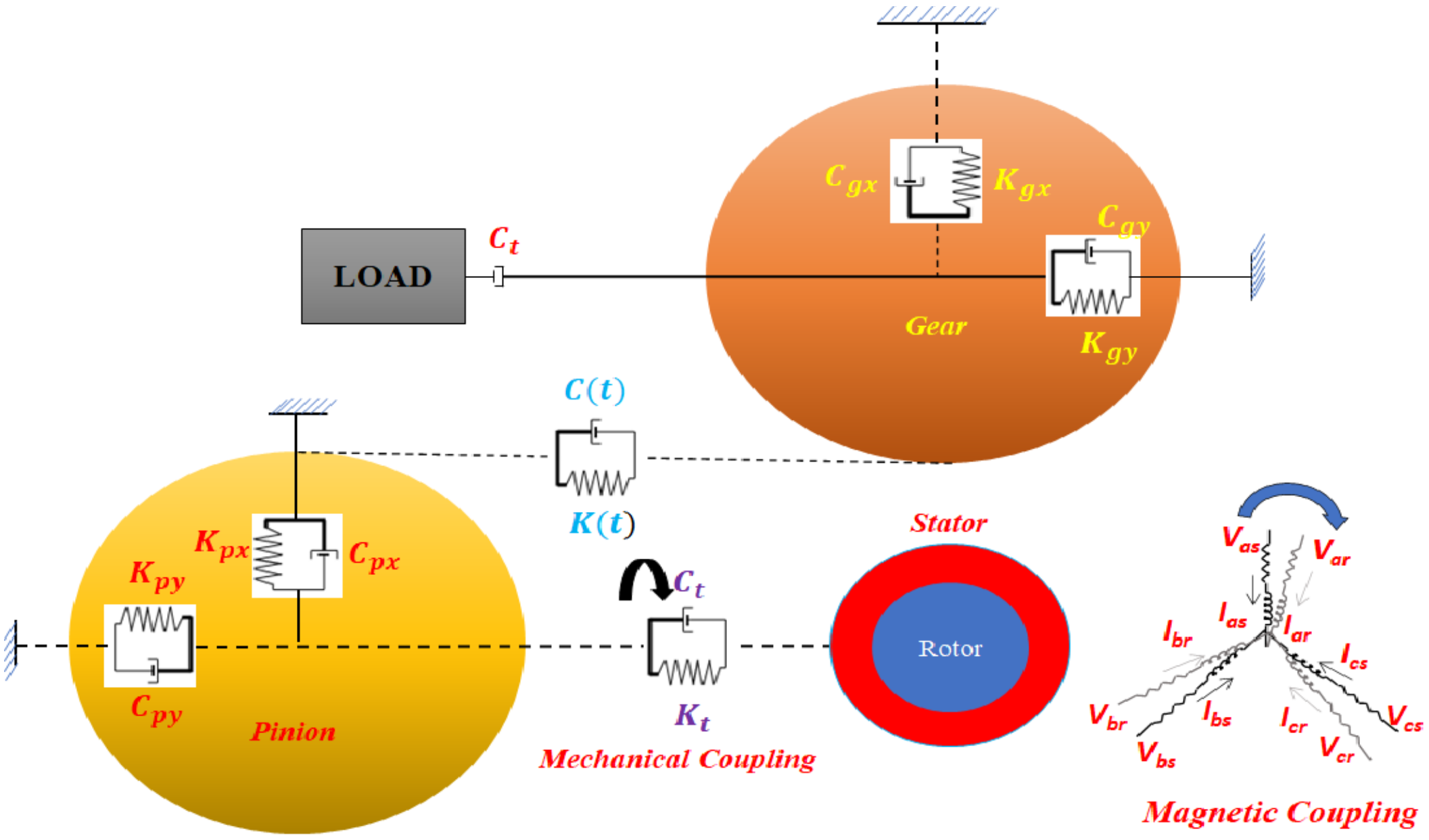

**Figure 2** Representation of Mathematical model of the electromechanical system [3]

### *4.2 IAM-TVMS*

The IAM-TVMS model developed in [4,7] was used to calculate the gear mesh stiffness of the gear pair for both stages. Analytically, this IAM-TVMS model has been improved by taking into account (1) the effects of misalignment among the tooth root and the base circle; (2) a precise changeover curve with an involute profile; (3) the effects of structural coupling and (4) the effects of nonlinear Hertzian contact. As a result, in the case of gear tooth 1, the total deformation along with the line of action can be calculated using the meshing force $F_1$, as follows:

$$\delta_1^i = \begin{cases} \delta_{t1}^i + \delta_{f1}^i, \\ \delta_{t1}^i + \delta_{f1}^i + \delta_{f12}^i, \\ \end{cases} \tag{25}$$

Similarly, in the case of gear tooth 2, the total deformation using meshing force $F_2$ along with the line of action may be calculated by:

$$\delta_2^i = \begin{cases} \delta_{t2}^i + \delta_{f2}^i, \\ \delta_{t2}^i + \delta_{f2}^i + \delta_{f21}^i, \end{cases} \tag{26}$$

Consequently, the mesh stiffness of tooth one and tooth two can be calculated by:

$$K_1^i = \frac{F_1}{\delta_1^i}, \quad K_2^i = \frac{F_2}{\delta_2^i}, \ (i = p, w) \tag{27}$$

Therefore, the total mesh stiffness $K(t)$ can be calculated as [4,7]:

$$K(t) = \begin{cases} \dfrac{1}{\left(\frac{1}{K_1^p} + \frac{1}{K_1^w} + \frac{1}{K_{h1}}\right)}, \\ \dfrac{1}{\left(\frac{1}{K_1^p} + \frac{1}{K_1^w} + \frac{1}{K_{h1}}\right)} + \dfrac{1}{\left(\frac{1}{K_2^p} + \frac{1}{K_2^w} + \frac{1}{K_{h2}}\right)}, \end{cases} \tag{28}$$

In Fig. 3, the consequences of the IAM-TVMS curve have been presented for single tooth as well as double tooth engagement, which demonstrate the influences of gear tooth root crack on the TVMS [3,4]. The IAM-TVMS curve tells us the gear tooth with healthy pair has a higher amount of TVMS than the cracked gear pair, and also, increasing gear tooth crack depths leads to the reduction in TVMS [3,4].

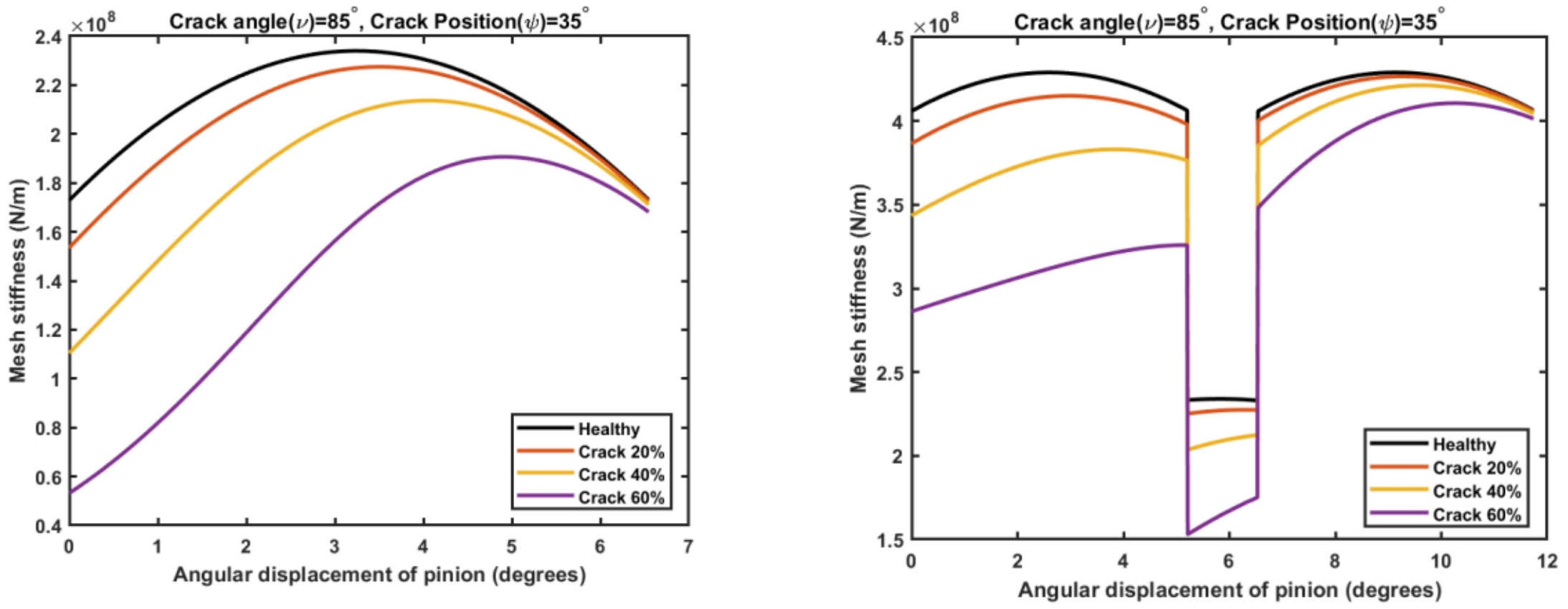

**Figure 3** TVMS graph of different tooth crack depths (a) single tooth engagement and (b) double tooth engagement [3,4].

For the dynamic simulation of the CEMG system, equation (8) to (13) has been written and solved as the state space equations in MATLAB employing the fourth order Runge-Kutta method. The system parameters of the CEMG system used in work can be found in [3]. The simulations ran for 4 seconds with the speed-load conditions of 25Hz-25lb and 25Hz-50lb. The dynamic response was taken based on the sampling frequency of 100KHz, leading to a total of 400000 vibration data points for 4 sec [3]. Also, a 10dB and -10dB White Gaussian noise has been added to make the dynamic response as accurate as environments. The simulations mainly took the *y*-directional vibration response of the CEMG system to demonstrate the consequence of different level gear tooth cracks, which were taken as 20%, 40%, and 60% gear tooth cracks, respectively. Fig. 4 shows the dynamic response of the CEMG system for 25Hz-25lb and 25Hz-50lb speed-load conditions with all gear tooth levels with -10dB and 10dB SNR.

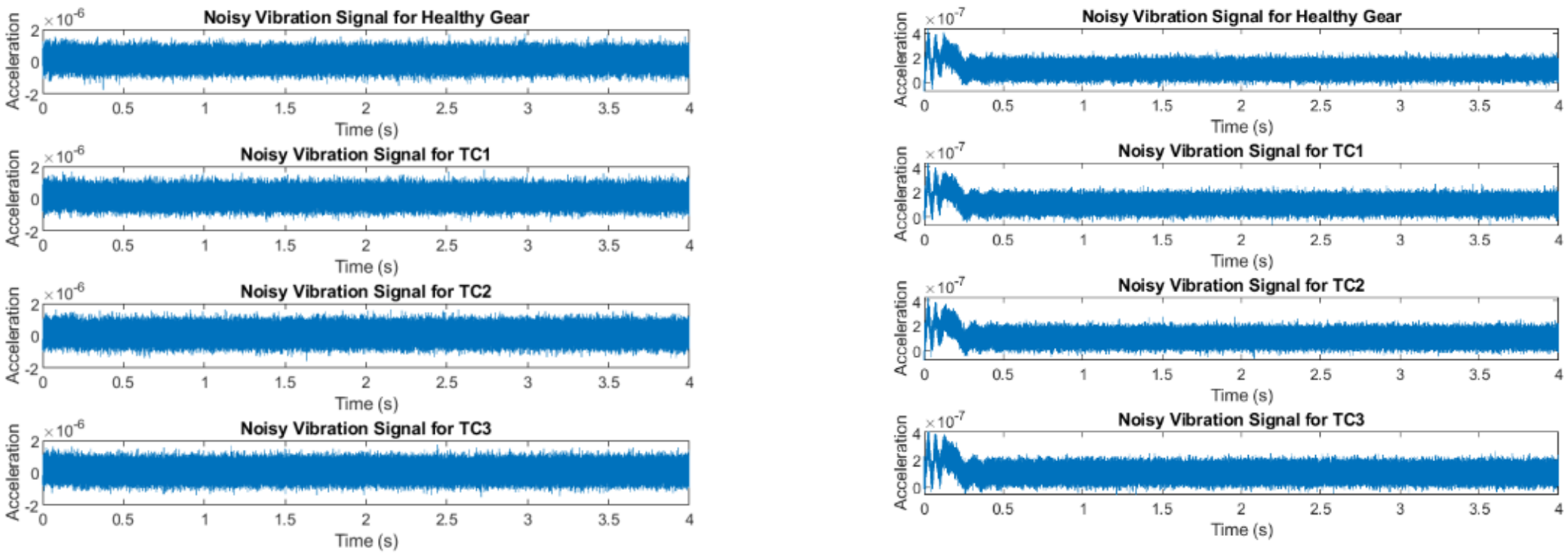

a b

**Figure 4** Simulated dynamic response with different levels of tooth crack depths with full data length [4] at 25Hz and 25lb with the addition of (a) -10dB & (b) 10dB SNR

### *4.3 VMD of Noisy Vibration Signal*

In section 2.1, the background theory of VMD has already been demonstrated. VMD filters out the non-stationary components of the noisy vibration response. VMD decomposes the vibration signal into multiple IMFs (VMFs), which helps to filter out fault transients from the noisy and variable vibration response [7,8,10]. Fig. 5 and Fig. 6 show the first three VMFs of five VMFs and the corresponding residual signal of the coupled electromechanical system with the addition of negative and positive white Gaussian noise at 25Hz and 25lb. All the individual VMFs consist of a unique center frequency [7,8,10]. Five VMFs with four gear tooth conditions have been considered for further investigations with TSA at 25Hz-25lb and 25Hz-50lb with the addition of -10dB and 10bB SNR.

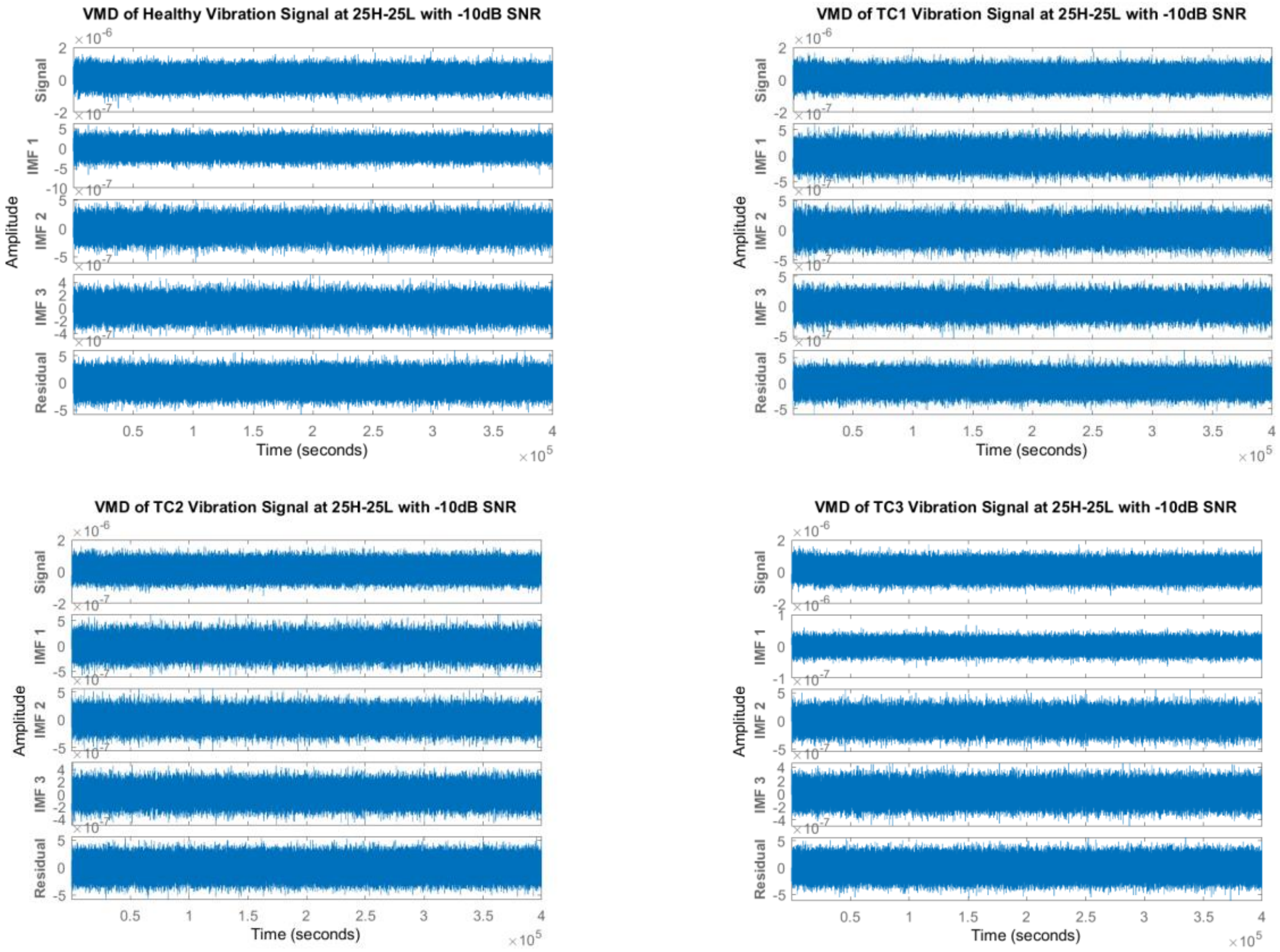

**Figure 5**. VMD of noisy vibration signals at 25hZ and 25lb with -10dB SNR

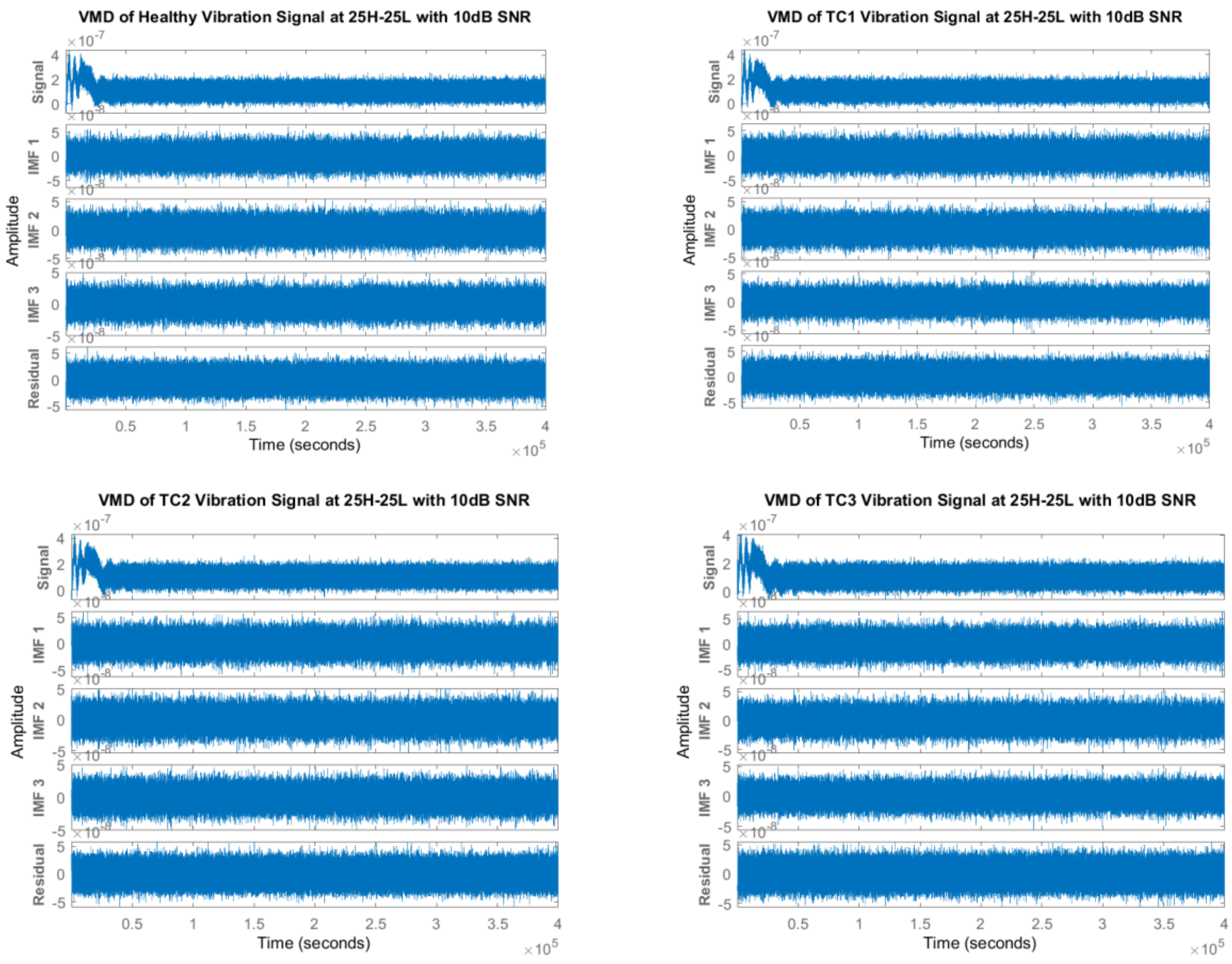

**Figure 6**. VMD of noisy vibration signals at 25hZ and 25lb with 10dB SNR

*4.4 TSA of VMF Vibration Signal*

TSA is a widely used signal preprocessing method that extracts the periodic waveforms of noisy vibration responses [7,11,12]. TSA is very convenient for vibration signals. TSA mainly allows the vibration response of a particular gear to be separated from other gears' vibrations and noise, which are not synchronous [7,11,12]. This section presents the TSA of VMF for all given conditions. Fig. 7 illustrates the TSA of noisy raw vibrations captured for healthy tooth conditions at 25Hz–25lb with additions of -10dB and 10dB SNR. It can be noted that vibration signals with -10dB contain more noise than 10dB SNR, which can be seen in Fig. 7. Similarly, Figs. 8 and 9 show TSA of the VMFs, which removes the effects of noise and asynchronous components from VMD vibration signals.

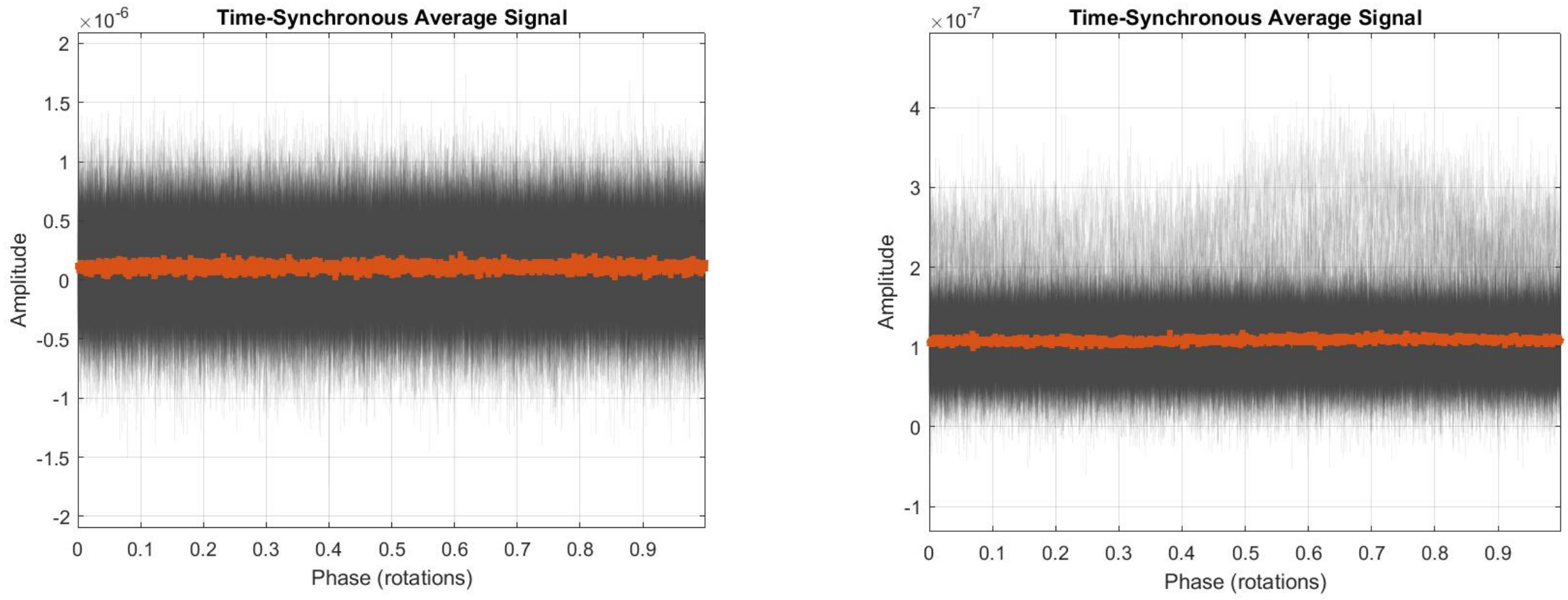

**Figure 7**. TSA signal of noisy vibration signal at 25Hz-25lb with -10dB SNR and 10dB SNR

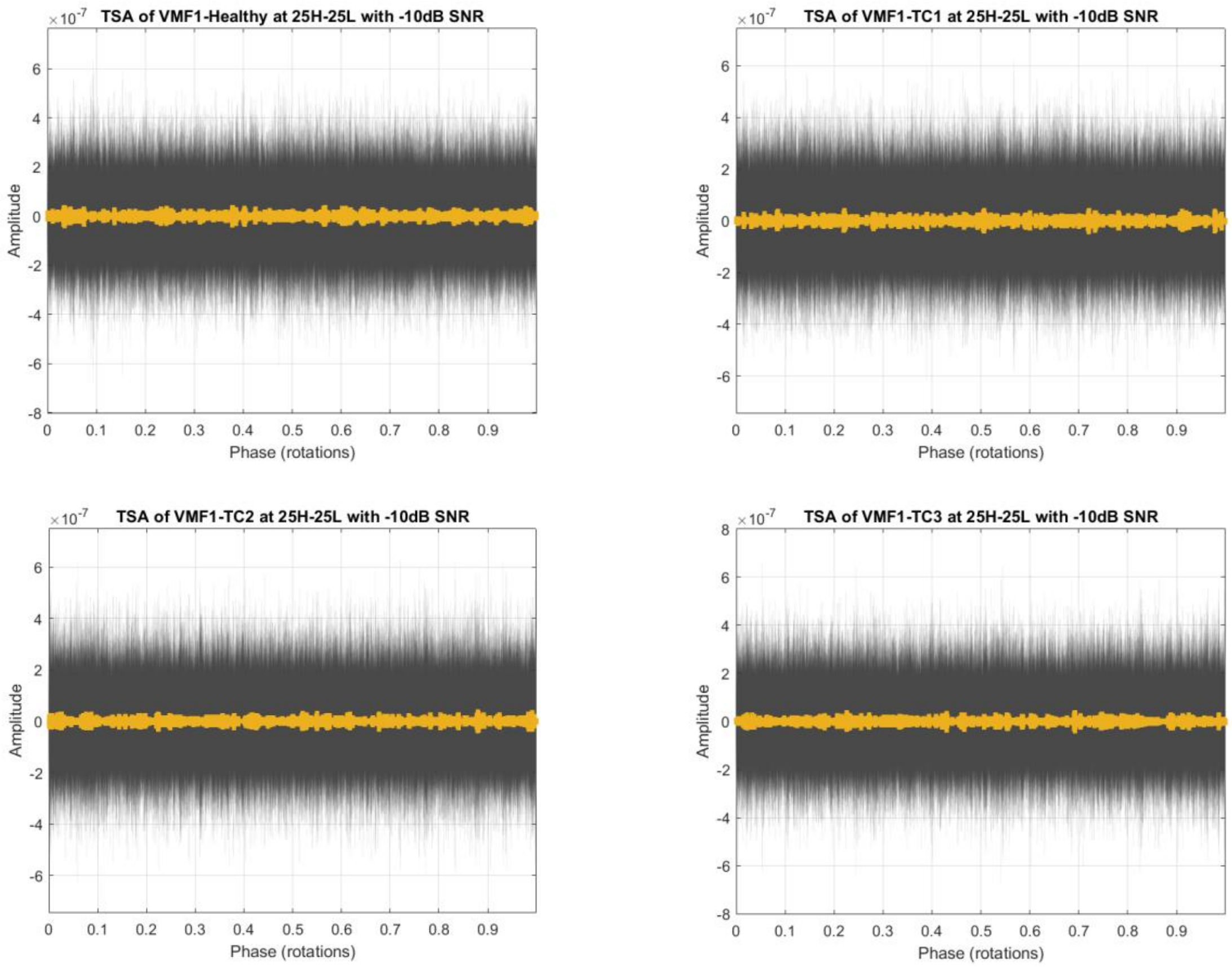

**Figure 8**. TSA of VMF1 at 25hZ and 25lb with -10dB SNR

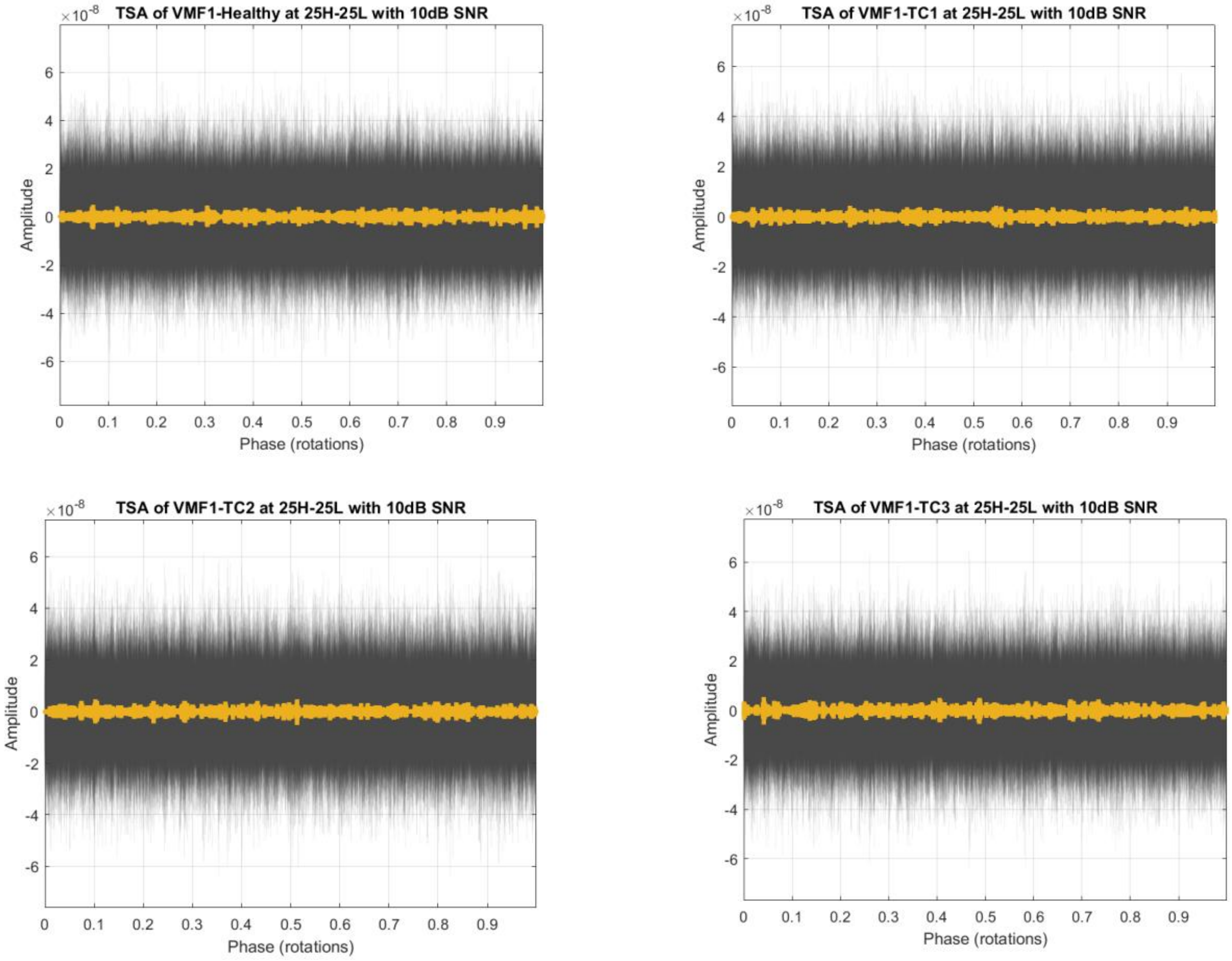

**Figure 9**. TSA of VMF1 at 25hZ and 25lb with 10dB SNR

### 4.5 *Lyapunov Exponent (LE) of VMF-TSA Vibration Signal*

The objective of using the LE algorithm with the combination of VMD-TSA is to determine the chaotic behaviors of a gear tooth fault crack [13-15,17]. In this work, the total number of data points filtered out non-stationary components of the vibration signal using VMD is 400000 for a 4-sec duration. Later, the same data points were considered to minimize noise and asynchronous components from VMFs, which became 4000 data points. This work calculated LE from the VMF-TSA vibration signal, which has 4000 data points. Fig. 10 and 11 show the LE of VMF-TSA with the addition of -10dB and 10dB White Gaussian noise in the case of 25Hz-25lb and 25Hz-50lb. Both figures denote that the LE of VMF1-TSA, VMF2-TSA, VMF3-TSA, and VMF4-TSA becomes positive LE. On the same note, it is crucial to observe that VMF5-TSA becomes negative LE in both figures. Although LE can be positive and negative, it depends on the input data on particular time durations [13-15,17]. In this case, the positive LE of VMF (1 to 4) -TSA indicates the vibration response of the gear tooth is nonlinear, and the negative LE indicates the generated vibration data of VMF5-TSA is in the stable or in the normal conditions/category [13-15,17]. Therefore, it can be concluded from Fig. 10 and 11 that VMF (1 to 4)-TSA has a more chaotic vibration than the VMF5-TSA because of the presence of positive LE [13-15,17].

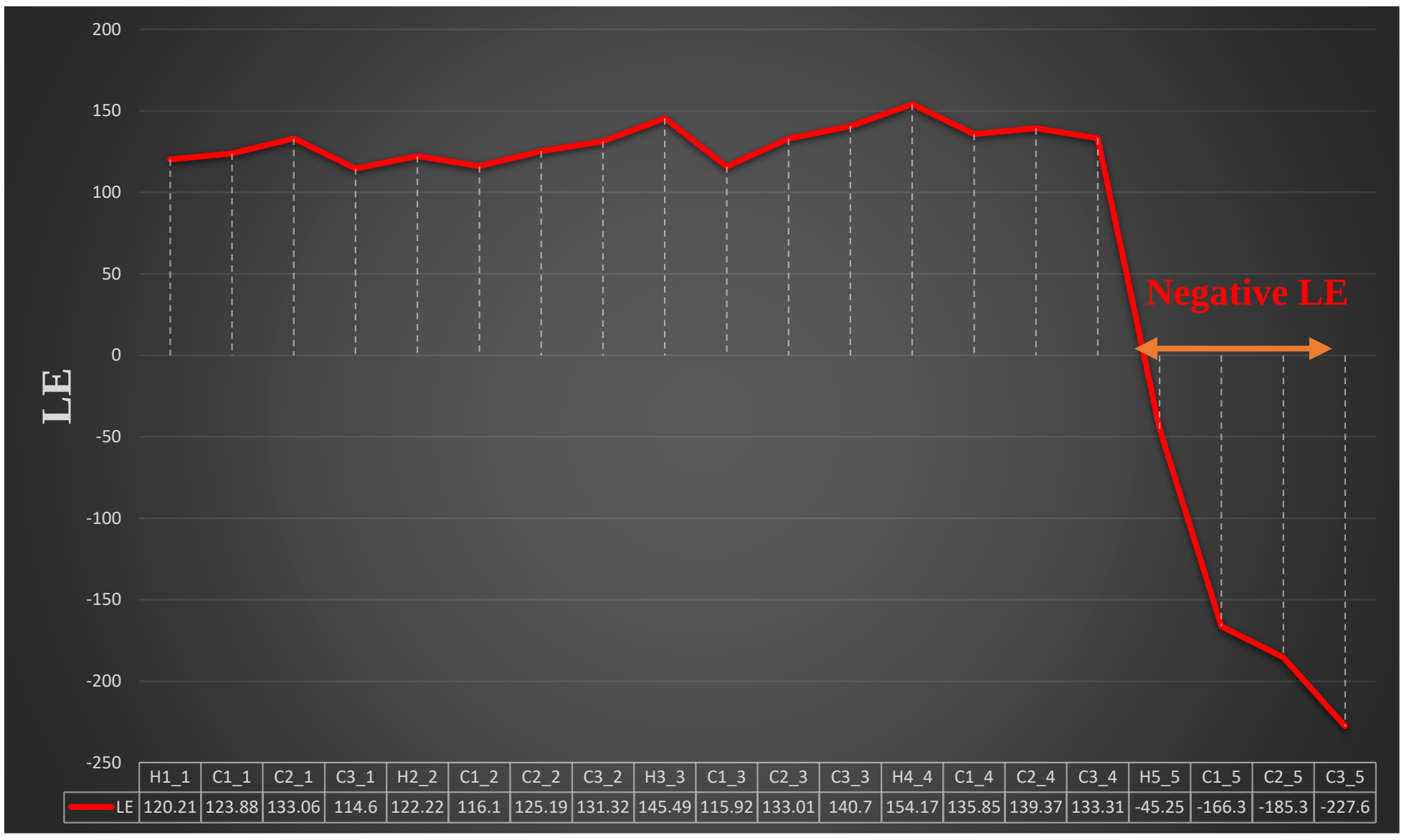

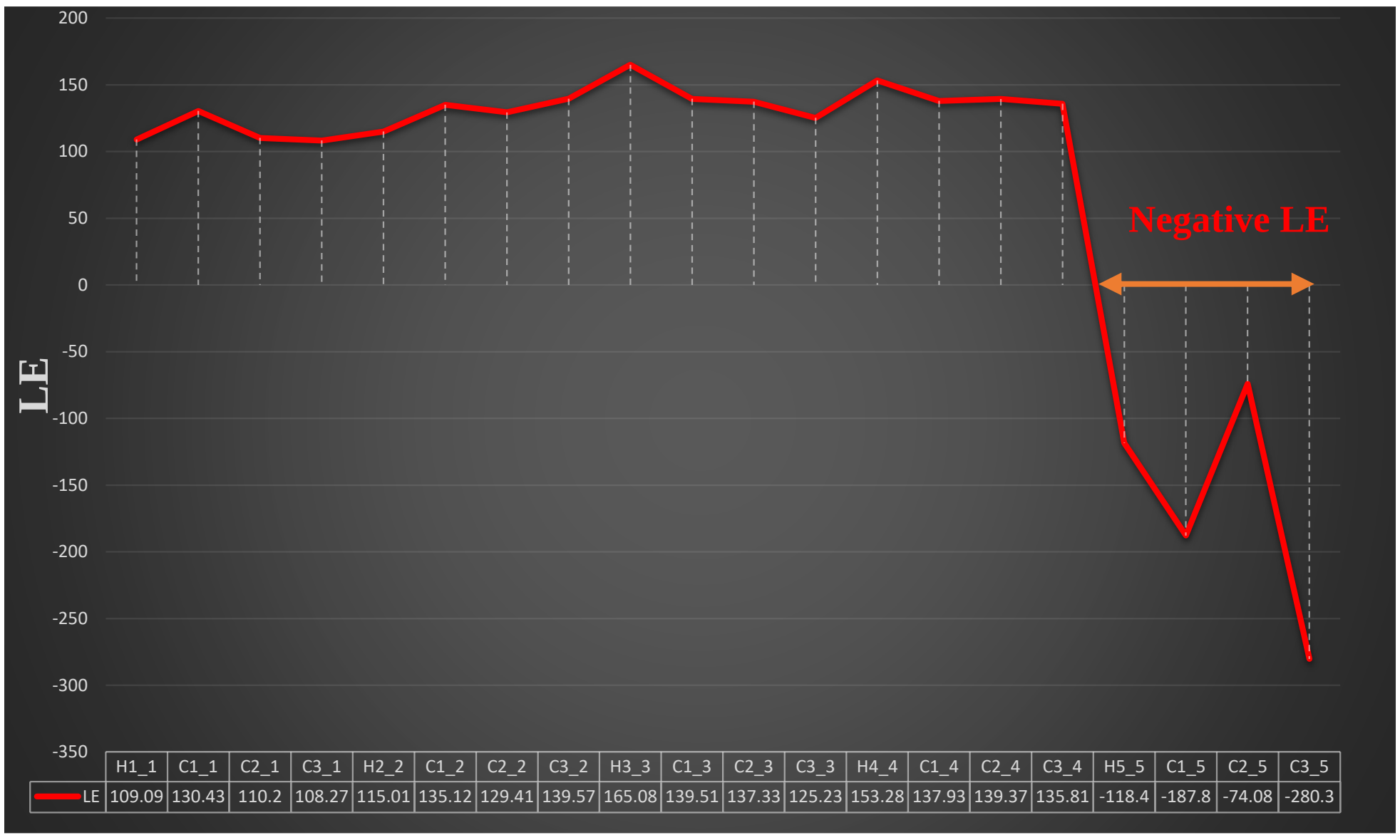

**Figure 10** Lyapunov exponent of various VMF-TSA various from H1_1, C1_1, C2_1, C3_1 for VMF1-TSA; H2_2, C1_2, C2_2, C3_2 for VMF2-TSA; H3_3, C1_3, C2_3, C3_3 for VMF3-TSA; H4_4, C1_4, C2_4, C3_4 for VMF4-TSA and H5_5, C1_5, C2_5, C3_5 for VMF5-TSA at 25Hz-25lb and 25Hz-50lb with the addition of -10dB White Gaussian noise

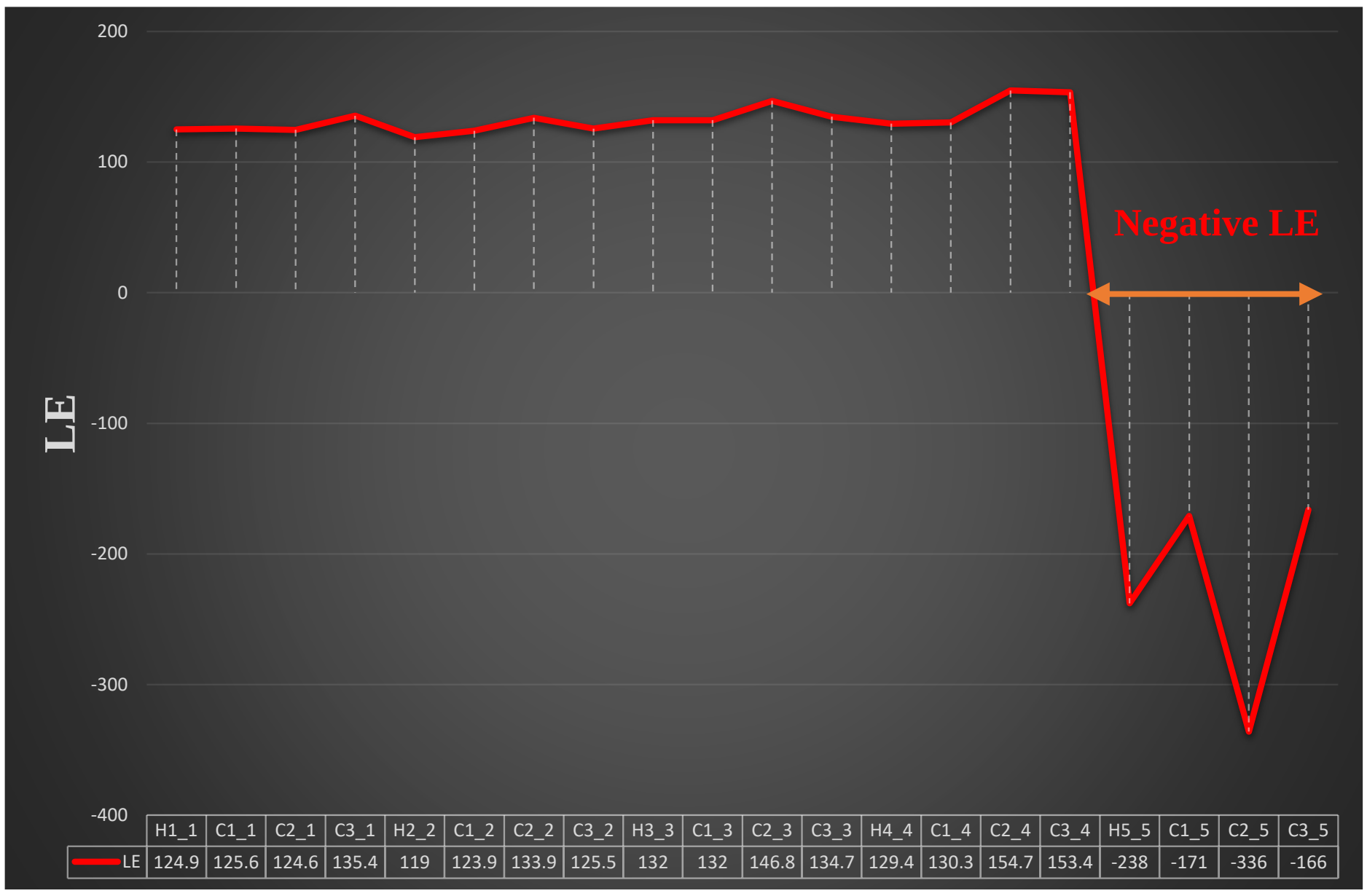

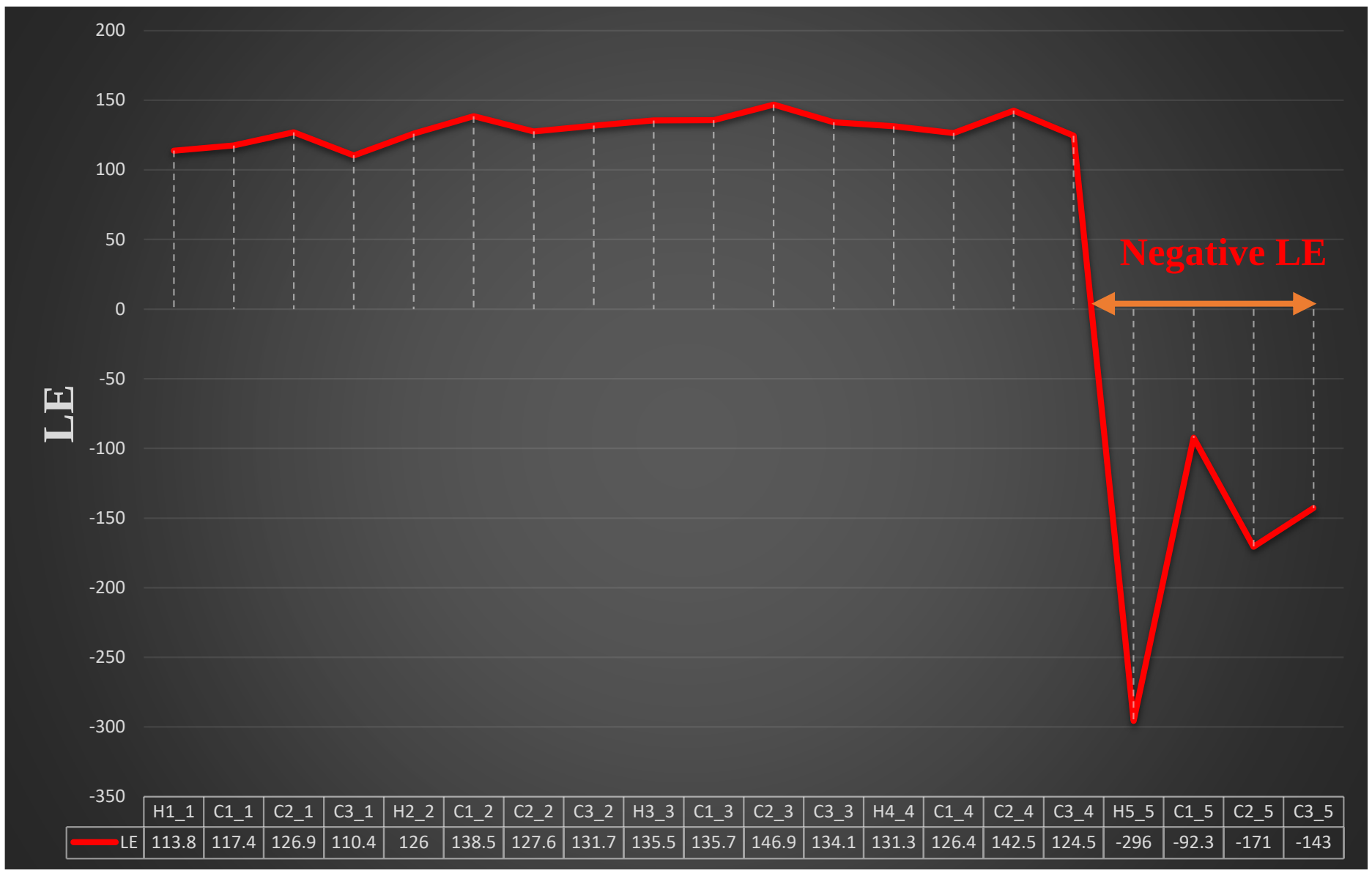

**Figure 11** Lyapunov exponent of various VMF-TSA various from H1_1, C1_1, C2_1, C3_1 for VMF1-TSA; H2_2, C1_2, C2_2, C3_2 for VMF2-TSA; H3_3, C1_3, C2_3, C3_3 for VMF3-TSA; H4_4, C1_4, C2_4, C3_4 for VMF4-TSA and H5_5, C1_5, C2_5, C3_5 for VMF5-TSA at 25Hz-25lb and 25Hz-50lb with the addition of 10dB White Gaussian noise.

### *4.6 Correlation Dimension (CD) of VMF-TSA Vibration Signal*

In Fig. 12, for the case of -10dB SNR, it can be observed that the VMF1-TSA of 25lb SNR, C3_1 shows a lower CD value than H1_1 and C1_1 & C2_1 show higher CD values than H1_1. In the case of 50 lb, C1_1 & C3_1 show the higher, and C2_2 shows the lower CD values of the H1_1. On the same note, for the case of VMF2_TSA in 25 lb, H2_2 shows higher CD values than the rest of the tooth crack conditions, and for 50 lb, all the tooth cracks show a continuous increase in the CD values. Similarly, in the case of 25lb_VMF3_TSA, which follows the parallel nature of VMF1-TSA, and for 50lb, VMF3-TSA also follows the exact nature as 50lb_VMF2_TSA. In the case of VMF4_TSA for 25 lb, C1_4 & C2_4 show the higher values of CD, and C1_4 shows the lower CD values. Parallelly, in the case of 50 lb, only C3_1 shows higher CD values than H4_4.

At last, 25lb_VMF5_TSA shows the H5_5 contains higher CD values than other tooth cracks, and in the case of 50lb_VMF5_TSA, which shows the continuous growth of CD values in the crack cases. In Fig. 13, in 10dB SNR, all of the CD shows a higher CD than the healthy case, except 25lb_VMF4_TSA, which shows the H4_4 has higher values than all of the crack cases, and also, in the case of VMF5_TSA of 25lb, H5_1 and C1_5 show quite similar CD values, and C2_5 and C3_3 show lower CD values than the others. Therefore, it can be concluded that the case of -10dB SNR with 25lb_VMF5_TSA and 10dB SNR with 25lb_VMF4_TSA only shows significantly improved CD values and that it also proved that increasing the tooth crack depths leads to a decrease in the CD values [13,16-18]. This indicates that a healthy gear tooth has higher chaotic complexity and less similarity than a cracked tooth.

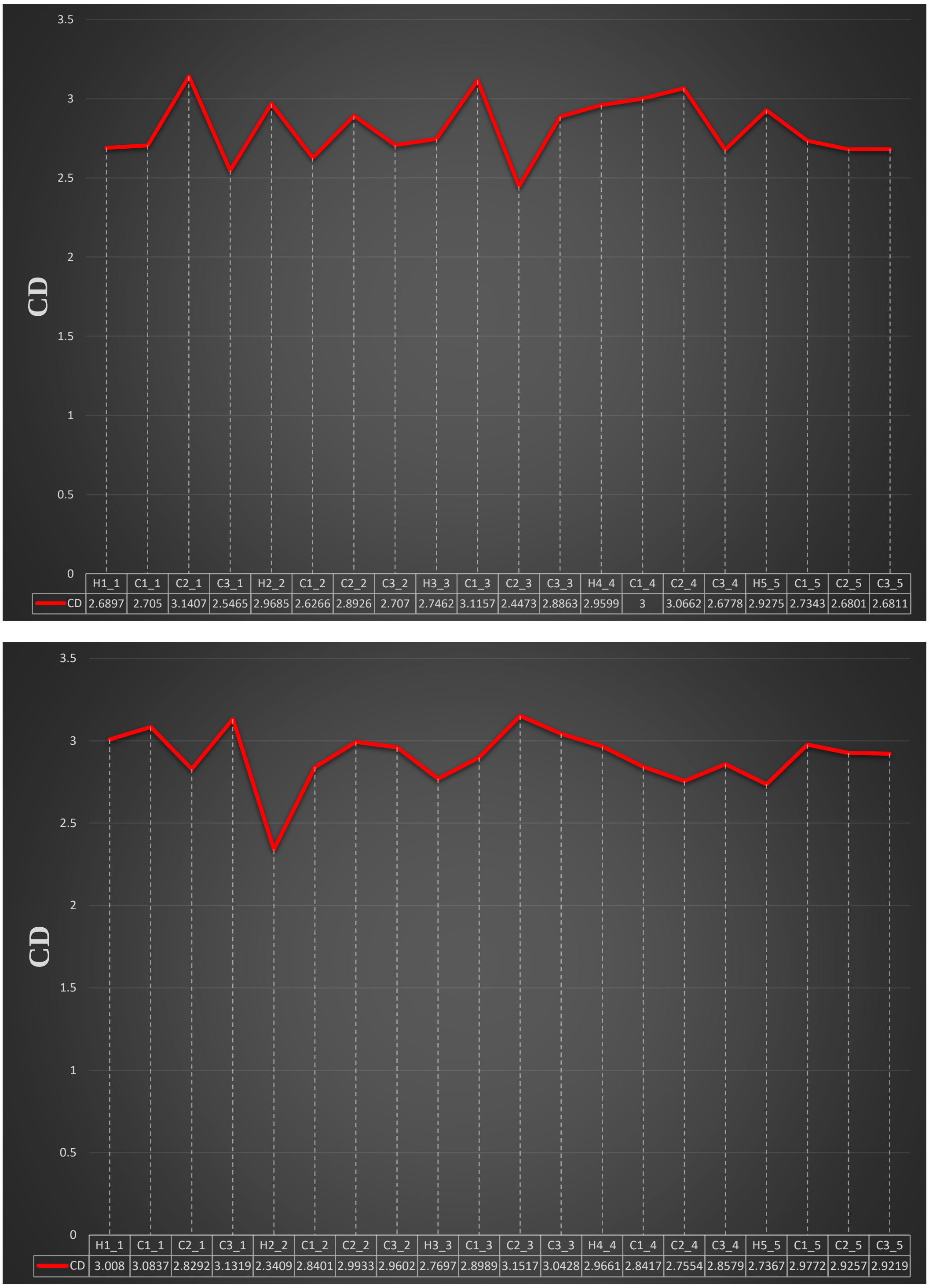

**Figure 12** Correlation dimension of various VMF-TSA various from H1_1, C1_1, C2_1, C3_1 for VMF1-TSA; H2_2, C1_2, C2_2, C3_2 for VMF2-TSA; H3_3, C1_3, C2_3, C3_3 for VMF3-TSA; H4_4, C1_4, C2_4, C3_4 for VMF4-TSA and H5_5, C1_5, C2_5, C3_5 for VMF5-TSA at 25Hz-25lb and 25Hz-50lb with the addition of -10dB White Gaussian noise.

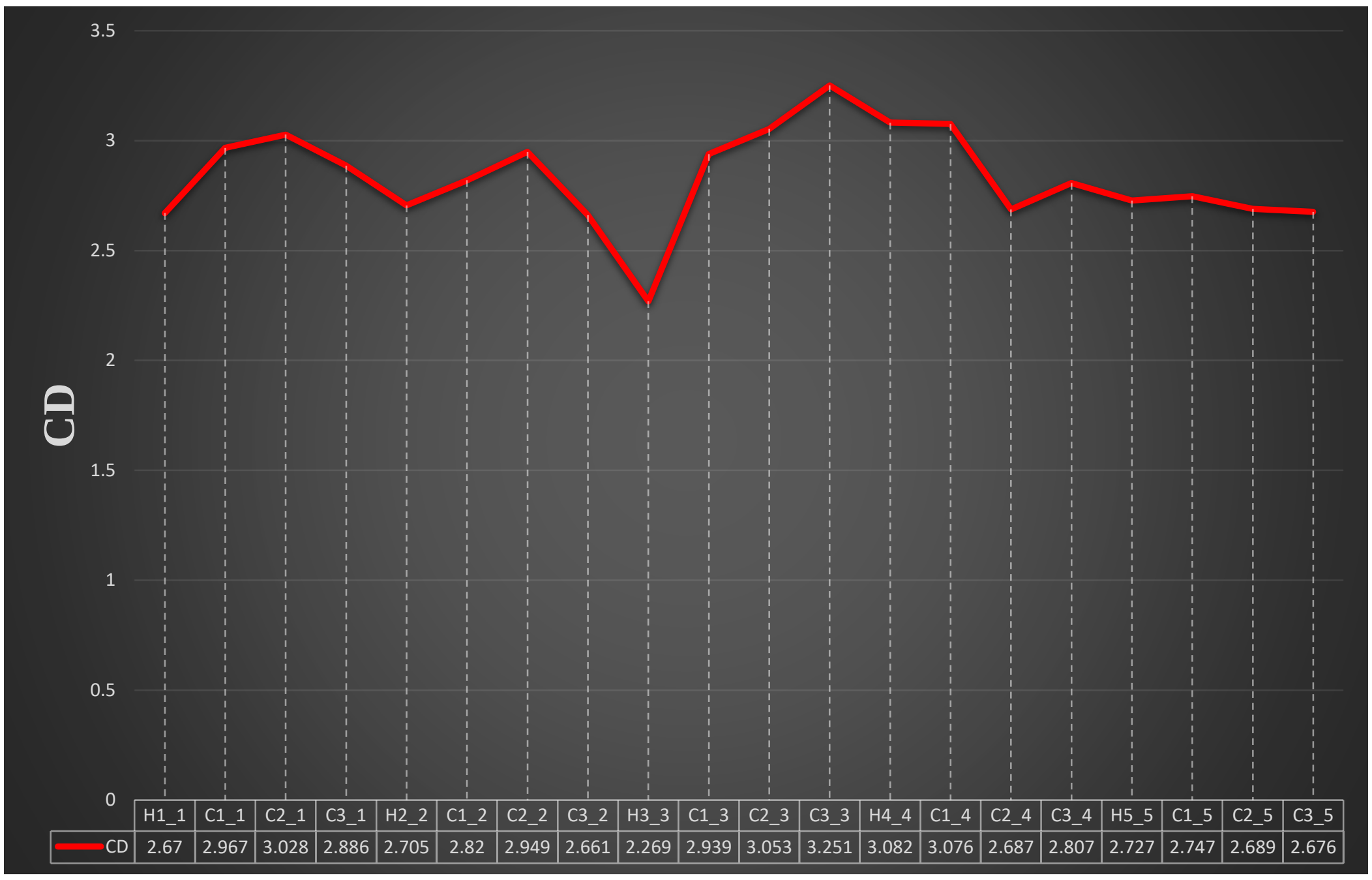

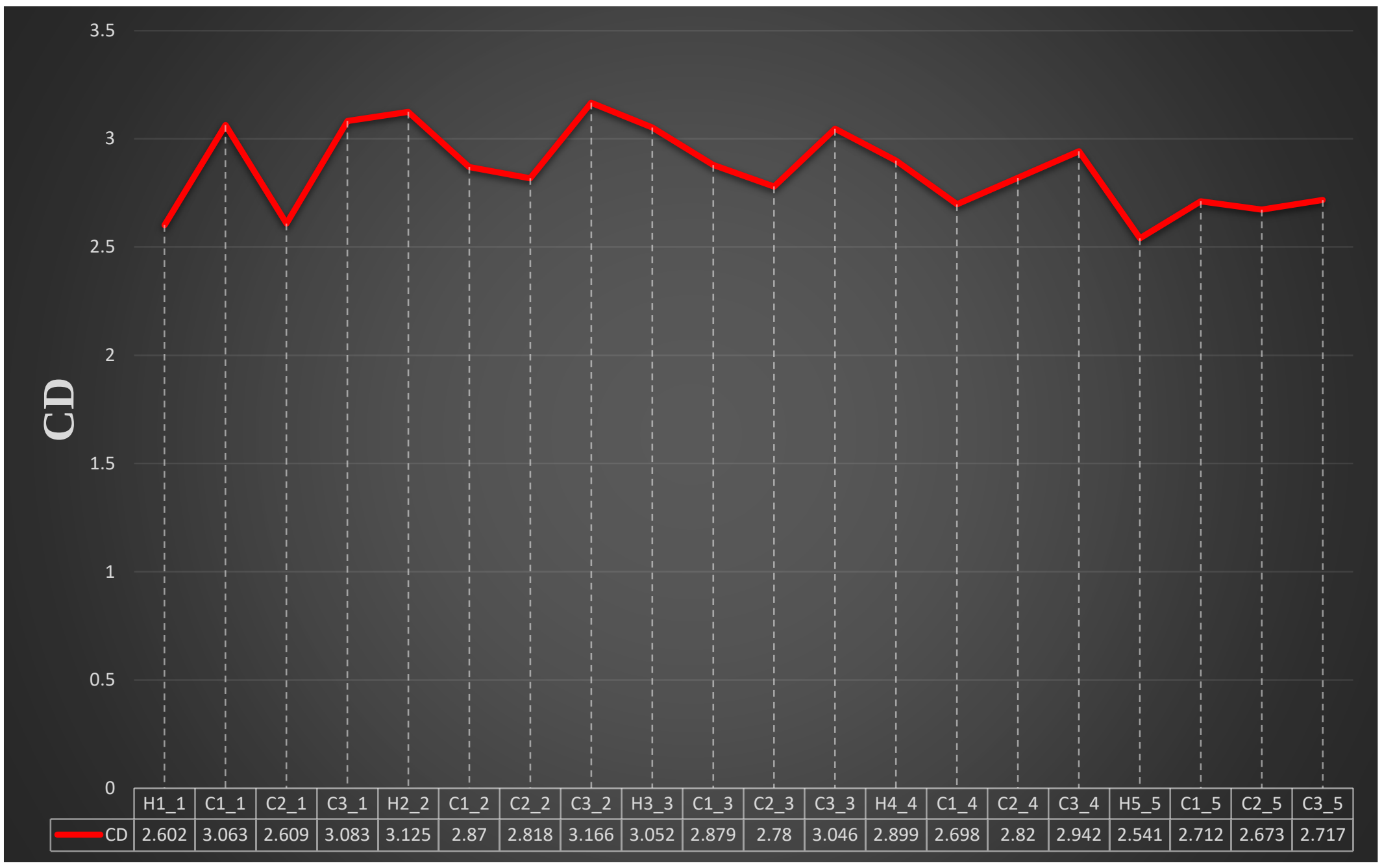

**Figure 13** Correlation dimension of various VMF-TSA various from H1_1, C1_1, C2_1, C3_1 for VMF1-TSA; H2_2, C1_2, C2_2, C3_2 for VMF2-TSA; H3_3, C1_3, C2_3, C3_3 for VMF3-TSA; H4_4, C1_4, C2_4, C3_4 for VMF4-TSA and H5_5, C1_5, C2_5, C3_5 for VMF5-TSA at 25Hz and 50lb with the addition of 10dB White Gaussian noise.

## 5. Conclusion

This paper presents an integrated approach to gear tooth cracks analysis based on the combined VMD-TSA. This paper incorporates the modified Lagrangian formulation to the mathematical modelling of the coupled electromechanical gearbox (CEMG) systems, which take Rayleigh's dissipative potential. This research also includes an analytically improved time-varying mesh stiffness with the modified Lagrangian formulation of coupled electromechanical (CEMG) systems. The combined VMD-TSA method has been used to study the dynamic response of the CEMG system with the addition of -10dB and 10dB SNR. The proposed method was applied and investigated based on the two speed-load cases, i.e., 25Hz-25lb and 25Hz-50lb, with four different gear tooth conditions, i.e., healthy (H), 20% tooth crack (C1), 40% tooth crack (C2), and 60% tooth crack (C3). The following essential points were observed during the development of the proposed methodology:

- This study was performed based on the 400000 data points used for VMD preprocessing techniques to filter out the nonstationarities from the noisy, complex, and non-stationary dynamic response of the CEMG system, which was later again preprocessed using TSA to remove the effects of noise and asynchronous components.
- VMF-TSA vibration signal used to calculate the chaotic features such as Lyapunov Exponent (LE) and Correlation Dimension (CD) on the basics of embedding dimension (m) = 3 and Lag (d) = 1.
- In the calculation of the Lyapunov Exponent (LE) of the CEMG system, VMF3-TSA and VMF4-TSA show that increasing the tooth crack depths leads to a decrease in the positive LE values for both cases of -10dB, i.e., 25Hz-25lb and 25Hz-50lb dynamic response.
- In the case of the Correlation Dimension (CD), -10dB SNR with 25lb_VMF5_TSA and 10dB SNR with 25lb_VMF4_TSA only shows significantly improved CD values, and it is also proved that increasing the tooth crack depths leads to a decrease in the CD values. This indicates that a healthy gear tooth has higher chaotic complexity and less similarity than a cracked tooth.
- Also, LE shows from Figs. 10 and 11 that VMF (1 to 4)-TSA has a more chaotic vibration than VMF5-TSA because of the presence of positive LE.

**Supplementary Results**

This is the graph of the enveloped spectrum of the VMF-TSA signal at 25Hz and 25lb with the addition of -10dB and 10dB SNR.

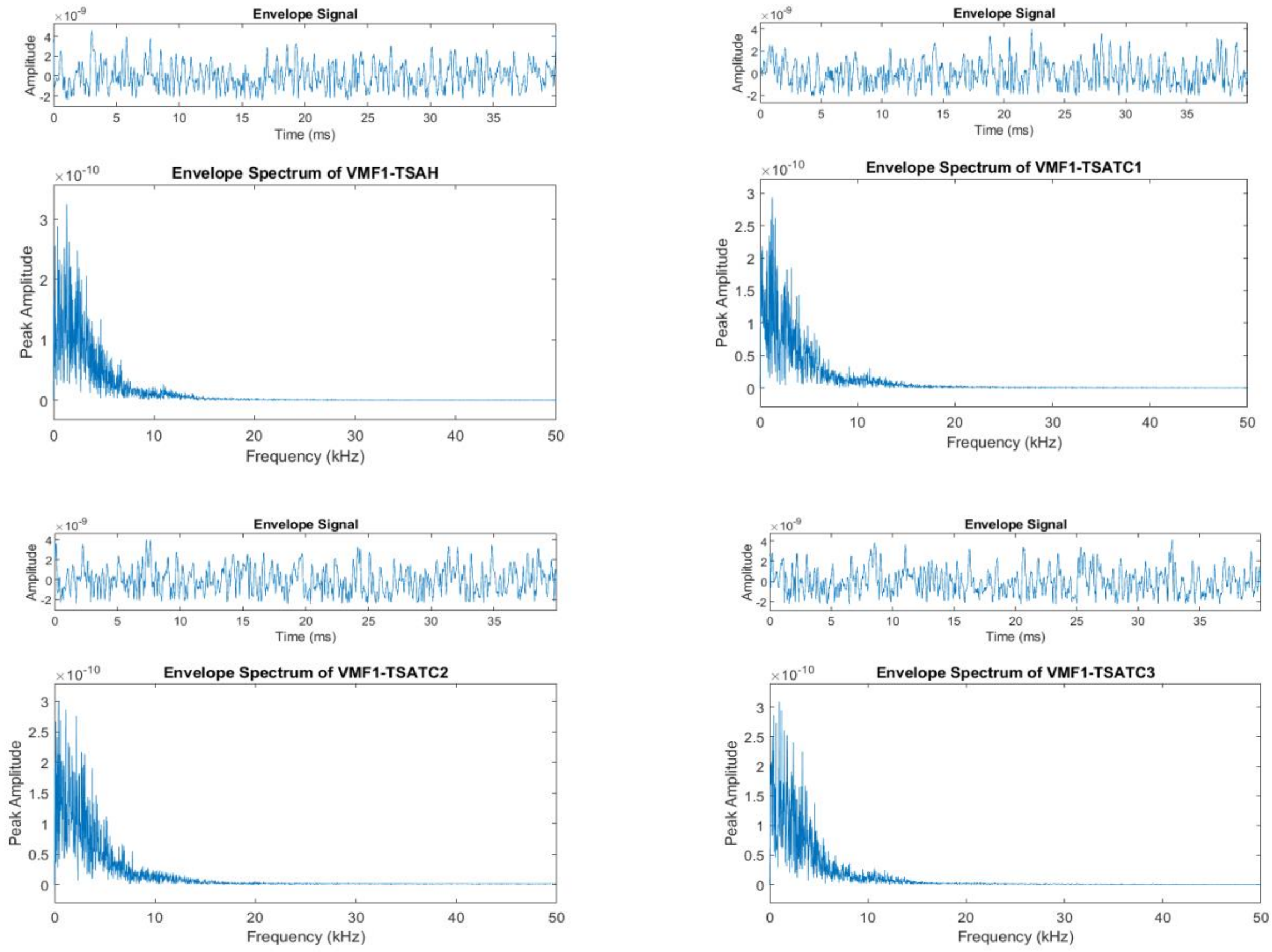

**Figure 1**. Envelope spectrum of VMF1-TSA at 25hZ and 25lb with -10dB SNR

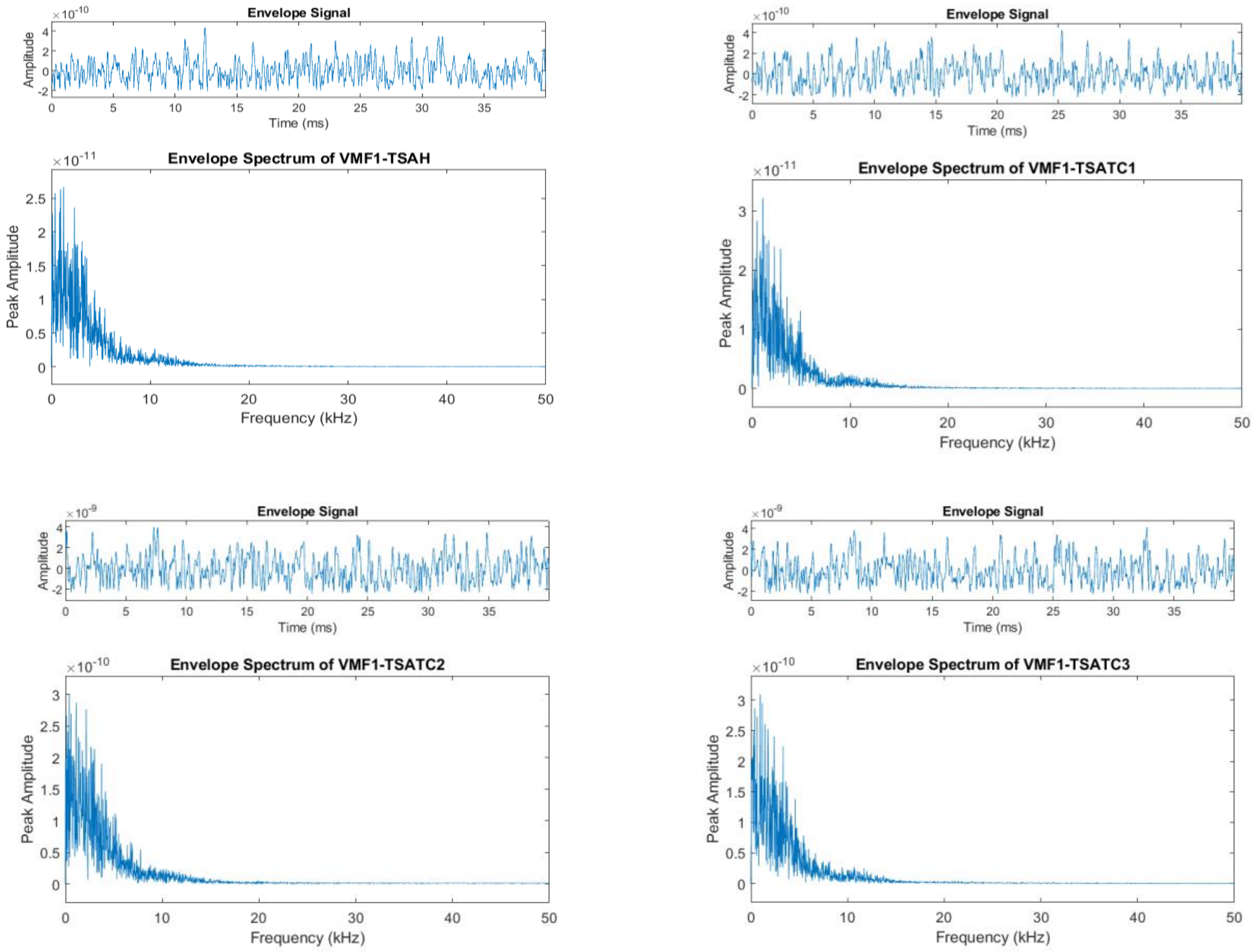

**Figure 2**. Envelope spectrum of VMF1-TSA at 25hZ and 25lb with 10dB SNR